\documentclass[onecolumn]{article}%
\usepackage{amsmath}
\usepackage{graphicx}%
\usepackage{amsfonts}%
\usepackage{amssymb}
\newtheorem{theorem}{Theorem}

\newtheorem{lemma}[theorem]{Lemma}

\newenvironment{proof}[1][Proof]{\textbf{#1.} }{\ \rule{0.5em}{0.5em}}
\begin{document}

\title{Quantum Lower Bound for the Collision Problem}
\author{Scott Aaronson\thanks{Computer Science Department, University of California,
Berkeley, CA 94720-1776. \ Email: \texttt{aaronson@cs.berkeley.edu}.
\ Supported in part by a National Science Foundation Graduate Fellowship and
by the Institute for Quantum Information at the California Institute of Technology.}}
\maketitle

\begin{abstract}
The collision problem is to decide whether a function $X:\left\{
1,\ldots,n\right\}  \rightarrow\left\{  1,\ldots,n\right\}  $\ is one-to-one
or two-to-one, given that one of these is the case. \ We show a lower bound of
$\Omega\left(  n^{1/5}\right)  $\ on the number of queries needed by a quantum
computer to solve this problem with bounded error probability. \ The best
known upper bound is $O\left(  n^{1/3}\right)  $, but obtaining any lower
bound better than $\Omega\left(  1\right)  $\ was an open problem since 1997.
\ Our proof uses the polynomial method augmented by some new ideas. \ We also
give a\ lower bound of $\Omega\left(  n^{1/7}\right)  $\ for the problem of
deciding whether two sets are equal or disjoint on a constant fraction of
elements. \ Finally we give implications of these results for quantum
complexity theory.

\end{abstract}

\section{Introduction}

The power of quantum computing has been intensively studied for a decade
\cite{bv,bbbv,grover,shor,bbcmw,ambainis,simon}. \ Apart from possible
applications---such as speeding up combinatorial search \cite{grover}\ and
breaking public-key cryptography \cite{shor}---a major motivation for this
work has been to better understand quantum theory itself. \ Thus, researchers
have tried to discover not just the capabilities of quantum computing but also
the limitations. \ This task is difficult, though; proving (for example) that
quantum computers cannot solve $NP$-complete problems in polynomial time would
imply $P\neq NP$.

A popular alternative is to study restricted models of computation, and
particularly the \textit{query model}, in which one counts only the number of
queries to the input, not the number of computational steps. \ An early result
of Bennett, Bernstein, Brassard, and Vazirani \cite{bbbv} showed that a
quantum computer needs $\Omega\left(  \sqrt{n}\right)  $\ queries to search a
list of $n$ items for one marked item. \ (This bound is tight, as evidenced by
Grover's algorithm \cite{grover}.) \ Subsequently, Beals et al. \cite{bbcmw},
Ambainis \cite{ambainis}, and others obtained lower bounds for many other problems.

But one problem, the collision problem, resisted attempts to prove a lower
bound \cite{bdhhmsw,ambainis}. \ Because of its simplicity, the problem was
widely considered a benchmark for our understanding of quantum query
complexity. \ The collision problem of size $n$, or $\operatorname*{Col}_{n}$,
is defined as follows. \ Let $X=x_{1}\ldots x_{n}$\ be a sequence of $n$
integers drawn from $\left\{  1,\ldots,n\right\}  $, with $n$ even. \ We are
guaranteed that either

\begin{enumerate}
\item[(1)] $X$ is one-to-one (that is, a permutation of $\left\{
1,\ldots,n\right\}  $), or

\item[(2)] $X$ is two-to-one (that is, each element of $\left\{
1,\ldots,n\right\}  $ appears in $X$ twice or not at all).
\end{enumerate}

The problem is to decide whether (1) or (2)\ holds.

We show that $Q_{2}\left(  \operatorname*{Col}_{n}\right)  =\Omega\left(
n^{1/5}\right)  $, where $Q_{2}$\ is bounded-error quantum query complexity as
defined by Beals et al. \cite{bbcmw}. \ Details of the oracle model are given
in Section \ref{prelim}. \ The best known upper bound, due to Brassard, H\o
yer, and Tapp \cite{bht},\ is $O\left(  n^{1/3}\right)  $; thus, our bound is
probably not tight. \ Previously, though, no lower bound better than the
trivial $\Omega\left(  1\right)  $\ bound was known. \ How great a speedup
quantum computers yield for the problem was apparently first asked by Rains
\cite{rains}.

Previous lower bound techniques failed for the problem because they depended
on a function's being sensitive to many disjoint changes to the input. \ For
example, Beals et al. \cite{bbcmw}\ showed that for all total Boolean
functions $f$, $Q_{2}\left(  f\right)  =\Omega\left(  \sqrt{\operatorname*{bs}%
\left(  f\right)  }\right)  $,\ where $\operatorname*{bs}\left(  f\right)
$\ is the block sensitivity, defined by Nisan \cite{nisan}\ to be, informally,
the maximum number of disjoint changes (to any particular input $X$) to which
$f$ is sensitive. \ In the case of the collision problem, though, every
one-to-one input differs from every two-to-one input in at least $n/2$ places,
so the block sensitivity is $O\left(  1\right)  $. \ Ambainis' adversary
method \cite{ambainis}, as currently formulated, faces a related obstacle.
\ In that method we consider the algorithm and input as a bipartite quantum
state, and upper-bound how much the \textit{entanglement} of the state can
increase via a single query. \ Yet under the simplest measures of
entanglement, the algorithm and input can become highly entangled after
$O\left(  1\right)  $\ queries, again because every one-to-one input is far
from every two-to-one input.

Our proof is an adaptation of the polynomial method, introduced to quantum
computing by Beals et al. \cite{bbcmw}. \ Their idea was to reduce questions
about quantum algorithms to easier questions about multivariate polynomials.
\ In particular, if a quantum algorithm makes $T$ queries, then its acceptance
probability is a polynomial over the input bits of degree at most $2T$. \ So
by showing that any polynomial approximating the desired output has high
degree, one obtains a lower bound on $T$.

To lower-bound the degree of a multivariate polynomial, a key technical trick
is to construct a related \textit{univariate} polynomial. \ Beals et al.
\cite{bbcmw}, using a lemma due to Minsky and Papert \cite{mp}, replace a
polynomial $p\left(  X\right)  $\ (where $X$ is a bit string) by $q\left(
\left|  X\right|  \right)  $\ (where $\left|  X\right|  $\ denotes the Hamming
weight of $X$), satisfying
\[
q\left(  k\right)  =\operatorname*{EX}\limits_{\left|  X\right|  =k}p\left(
X\right)  \
\]
and $\deg\left(  q\right)  \leq\deg\left(  p\right)  $.

We construct the univariate polynomial in a different way. \ We consider a
uniform distribution over $k$-to-one inputs, where $k$ might be greater than
$2$. \ Even though the problem is to distinguish $k=1$\ from $k=2$, the
acceptance probability must lie between $0$ and $1$ for all $k$, and that is a
surprisingly strong constraint. \ We show that the acceptance probability is
\textit{close }to a univariate polynomial in $k$ of degree at most $2T$. \ We
then obtain a lower bound by generalizing a classical approximation theory
result of Ehlich and Zeller \cite{ez}\ and Rivlin and Cheney \cite{rc}. \ Much
of the proof deals with the complication that $k$ does not divide $n$ in general.

Shi \cite{shi}\ has recently improved our method to obtain a lower bound of
$\Omega\left(  n^{1/4}\right)  $ for the collision problem.

The paper is organized as follows. \ Section \ref{motivation}\ motivates the
collision lower bound within quantum computing, pointing out connections to
collision-resistant hash functions, the nonabelian hidden subgroup problem,
and information erasure. \ Section \ref{prelim}\ gives technical
preliminaries, Section \ref{bivar}\ proves the crucial fact that the
acceptance probability is ``almost''\ a univariate polynomial, and Section
\ref{lb}\ completes the lower bound argument. \ In Appendix \ref{setcomp}\ we
show a lower bound of $\Omega\left(  n^{1/7}\right)  $\ for the \textit{set
comparison problem}, a variant of the collision problem that is needed for the
application to information erasure.

\section{Motivation\label{motivation}}

The most immediate implication of the collision lower bound is that certain
problems, notably breaking cryptographic hash functions, are not in $BQP$
relative to an oracle. \ A second implication is that a nonstandard quantum
oracle model proposed by Kashefi et al. \cite{kashefi}\ is exponentially more
powerful than the usual oracle model. \ A third implication, in our view the
most interesting one, concerns the computational power of so-called
\textit{dynamical quantum theories}. \ That implication will be discussed in
detail in another paper.

\subsection{Oracle Hardness Results}

The original motivation for the collision problem was to model
\textit{(strongly) collision-resistant hash functions} in cryptography.
\ There is a large literature on collision-resistant hashing; see
\cite{damgard,bsp}\ for example. \ When building secure digital signature
schemes, it is useful to have a family of hash functions $\left\{
H_{i}\right\}  $, such that finding a distinct $\left(  x,y\right)  $ pair
with $H_{i}\left(  x\right)  =H_{i}\left(  y\right)  $\ is computationally
intractable. \ A quantum algorithm for finding collisions using $O\left(
\operatorname*{polylog}\left(  n\right)  \right)  $ queries\ would render
\textit{all} hash functions insecure against quantum attack in this sense.
\ (Shor's algorithm \cite{shor}\ already renders hash functions based on
modular arithmetic insecure.) \ Our result indicates that collision-resistant
hashing might still be possible in a quantum setting.

The collision problem also models the \textit{nonabelian hidden subgroup
problem}, of which graph isomorphism is a special case. \ Given a group $G$
and subgroup $H\leq G$, suppose we have oracle access to a function
$f:G\rightarrow\mathbb{N}$\ such that for all $g_{1},g_{2}\in G$, $f\left(
g_{1}\right)  =f\left(  g_{2}\right)  $ if and only if $g_{1}$\ and $g_{2}%
$\ belong to the same coset of $H$. \ Is there then an efficient quantum
algorithm to determine $H$? \ If $G$ is abelian, the work of Simon
\cite{simon}\ and Shor \cite{shor}\ implies an affirmative answer. \ If $G$ is
nonabelian, though, efficient quantum algorithms are known only for special
cases \cite{eh,gsvv}. \ An $O\left(  \operatorname*{polylog}\left(  n\right)
\right)  $-query algorithm for the collision problem would yield a
polynomial-time algorithm to distinguish $\left|  H\right|  =1$\ from $\left|
H\right|  =2$, which does not exploit the group structure at all. \ Our result
implies that no such algorithm exists.

\subsection{Information Erasure\label{erasure}}

Let $f:\left\{  0,1\right\}  ^{n}\rightarrow\left\{  0,1\right\}  ^{m}$ with
$m\geq n$\ be a one-to-one function. \ Then we can consider two kinds of
quantum oracle for $f$:

\begin{enumerate}
\item[(A)] a \textit{standard oracle}, one that maps $\left|  x\right\rangle
\left|  z\right\rangle $ to $\left|  x\right\rangle \left|  z\oplus f\left(
x\right)  \right\rangle $, or

\item[(B)] an \textit{erasing oracle} (as recently proposed by Kashefi et al.
\cite{kashefi}), which maps\ $\left|  x\right\rangle $ to $\left|  f\left(
x\right)  \right\rangle $, in effect ``erasing'' $\left|  x\right\rangle $.
\end{enumerate}

Intuitively erasing oracles seem at least as strong as standard ones, though
it is not clear how to simulate the latter with the former without also having
access to an oracle that maps $\left|  y\right\rangle $ to $\left|
f^{-1}\left(  y\right)  \right\rangle $. \ The question that concerns us here
is whether erasing oracles are \textit{more} useful than standard ones for
some problems. \ One-way functions provide a clue: if $f$ is one-way, then (by
assumption) $\left|  x\right\rangle \left|  f\left(  x\right)  \right\rangle
$\ can be computed efficiently, but if $\left|  f\left(  x\right)
\right\rangle $\ could be computed efficiently given $\left|  x\right\rangle
$\ then so could $\left|  x\right\rangle $\ given $\left|  f\left(  x\right)
\right\rangle $, and hence $f$ could be inverted. \ But can we find, for some
problem, an exponential gap between query complexity given a standard oracle
and query complexity given an erasing oracle?

In Appendix \ref{setcomp}\ we extend the collision lower bound to show an
affirmative answer. \ Define the \textit{set comparison problem} of size $n$,
or $\operatorname*{SetComp}_{n}$, as follows. \ We are given as input two
sequences, $X=x_{1}\ldots x_{n}$ and $Y=y_{1}\ldots y_{n}$, such that for each
$i$, $x_{i},y_{i}\in\left\{  1,\ldots,2n\right\}  $. \ A query has the form
$\left(  b,i\right)  $, where $b\in\left\{  0,1\right\}  \ $and $i\in\left\{
1,\ldots,n\right\}  $, and produces as output $\left(  0,x_{i}\right)  $\ if
$b=0$\ and $\left(  1,y_{i}\right)  $\ if $b=1$. $\ $Sequences $X$ and $Y$ are
both one-to-one; that is, $x_{i}\neq x_{j}$\ and $y_{i}\neq y_{j}$\ for all
$i\neq j$. \ We are furthermore guaranteed that either

\begin{enumerate}
\item[(1)] $X$ and $Y$ are equal as sets (that is, $\left\{  x_{1}%
,\ldots,x_{n}\right\}  =\left\{  y_{1},\ldots,y_{n}\right\}  $) or

\item[(2)] $X$ and $Y$ are far as sets (that is, $\left|  \left\{
x_{1},\ldots,x_{n}\right\}  \cup\left\{  y_{1},\ldots,y_{n}\right\}  \right|
\geq1.1n$).
\end{enumerate}

As before the problem is to decide whether (1) or (2) holds.

This problem can be solved with high probability in a constant number of
queries using an erasing oracle, by using a trick similar to that of Watrous
\cite{watrous}\ for verifying group non-membership. \ First, using the oracle,
we prepare the uniform superposition%
\[
\frac{1}{\sqrt{2n}}\sum_{i\in\left\{  1,\ldots,n\right\}  }\left(  \left|
0\right\rangle \left|  x_{i}\right\rangle +\left|  1\right\rangle \left|
y_{i}\right\rangle \right)  \text{.}%
\]
We then apply a Hadamard gate to the first register, and finally we measure
the first register. \ If $X$ and $Y$ are equal as sets, then interference
occurs between every $\left(  \left|  0\right\rangle \left|  z\right\rangle
,\left|  1\right\rangle \left|  z\right\rangle \right)  $\ pair and we observe
$\left|  0\right\rangle $\ with certainty. \ But if $X$ and $Y$ are far as
sets, then basis states $\left|  b\right\rangle \left|  z\right\rangle $\ with
no matching $\left|  1-b\right\rangle \left|  z\right\rangle $\ have
probability weight at least $1/10$, and hence we observe $\left|
1\right\rangle $\ with probability at least $1/20$.

In Appendix \ref{setcomp}\ we show that $Q_{2}\left(  \operatorname*{SetComp}%
_{n}\right)  =\Omega\left(  n^{1/7}\right)  $; that is, no efficient quantum
algorithm using a standard oracle exists for this problem.

\section{Preliminaries\label{prelim}}

Let $A$ be a quantum query algorithm. \ A basis state of $A$ is written
$\left|  \Psi,i,z\right\rangle $. \ Then a query replaces each $\left|
\Psi,i,z\right\rangle $ by $\left|  \Psi\oplus x_{i},i,z\right\rangle $, where
$x_{i}$\ is exclusive-OR'ed into some specified location of $\Psi$\ (which we
cannot assume to be all $0$'s). \ We assume without loss of generality that
every basis state queries at every step. \ Between queries, the algorithm can
perform any unitary operation that does not depend on the input. \ At the end
$z$ is measured in the standard basis;\ if $z=1$\ the algorithm returns
`one-to-one'\ and if $z=2$\ it returns `two-to-one.' \ The total number of
queries is denoted $T$. \ Also, we assume for simplicity that all amplitudes
are real; this restriction is without loss of generality \cite{bv}.

Let $\alpha_{X,\Psi,i,z}^{\left(  t\right)  }$\ be the amplitude of basis
state $\left|  \Psi,i,z\right\rangle $\ after $t$ queries when the input is
$X$. \ Also, let $\Delta\left(  x_{i},h\right)  =1$\ if $x_{i}=h$, and
$\Delta\left(  x_{i},h\right)  =0$ if $x_{i}\neq h$. \ Let $P\left(  X\right)
$ be the probability that $A$ returns $z=2$ when the input is $X$. \ Then we
obtain a simple variant of the main lemma of Beals et al. \cite{bbcmw}.

\begin{lemma}
\label{poly}$P\left(  X\right)  $ is a multilinear polynomial of degree at
most $2T$ over the $\Delta\left(  x_{i},h\right)  $.
\end{lemma}

\begin{proof}
We show, by induction on $t$, that for all basis states $\left|
\Psi,i,z\right\rangle $,\ $\alpha_{X,\Psi,i,z}^{\left(  t\right)  }$\ is a
multilinear polynomial of degree at most $t$ over the $\Delta\left(
x_{i},h\right)  $. \ Since $P\left(  X\right)  $\ is a sum of squares of
$\alpha_{X,\Psi,i,z}^{\left(  t\right)  }$, the lemma follows.

The base case ($t=0$) holds since, before making any queries, each
$\alpha_{X,\Psi,i,z}^{\left(  0\right)  }$\ is a degree-$0$ polynomial over
the $\Delta\left(  x_{i},h\right)  $. \ A unitary transformation on the
algorithm part replaces each $\alpha_{X,\Psi,i,z}^{\left(  t\right)  }$\ by a
linear combination of $\alpha_{X,\Psi,i,z}^{\left(  t\right)  }$, and hence
cannot increase the degree. \ Suppose the lemma holds prior to the $t^{th}%
$\ query. \ Then%
\[
\alpha_{X,\Psi,i,z}^{\left(  t+1\right)  }=\sum_{1\leq h\leq n}\alpha
_{X,\Psi\oplus h,i,z}^{\left(  t\right)  }\Delta\left(  x_{i},h\right)  ,
\]
and we are done.
\end{proof}

A remark on notation: we sometimes use brackets ($a_{b\left[  c\right]  }$)
rather than nested subscripts ($a_{b_{c}}$).

\section{Reduction to Bivariate Polynomial\label{bivar}}

Call the point $\left(  g,N\right)  \in\Re^{2}$\ an $\left(  n,T\right)
$-\textit{quasilattice point} if and only if

\begin{enumerate}
\item[(1)] $g$ and $N$ are integers, with $g$ dividing $N$,

\item[(2)] $1\leq g\leq\sqrt{n}$,

\item[(3)] $n\leq N\leq n+n/\left(  10T\right)  $, and

\item[(4)] if $g=1$\ then $N=n$.
\end{enumerate}

For quasilattice point $\left(  g,N\right)  $,\ define $\mathcal{D}_{n}\left(
g,N\right)  $\ to be the uniform distribution over all size-$n$ subfunctions
of $g$-1\ functions having domain $\left\{  1,\ldots,N\right\}  $\ and range a
subset of $\left\{  1,\ldots,n\right\}  $. \ More precisely: to draw an $X$
from $\mathcal{D}_{n}\left(  g,N\right)  $, we first choose a set
$S\subseteq\left\{  1,\ldots,n\right\}  $\ with $\left|  S\right|  =N/g\leq
n$\ uniformly at random. \ We then choose a $g$-1 function $\widehat
{X}=\widehat{x}_{1}\ldots\widehat{x}_{N}$ from $\left\{  1,\ldots,N\right\}
$\ to $S$ uniformly at random. \ Finally we let $x_{i}=\widehat{x}_{i}$\ for
each $1\leq i\leq n$.

Let $P\left(  g,N\right)  $\ be the probability that algorithm $A$\ returns
$z=2$\ when the input is chosen from $\mathcal{D}_{n}\left(  g,N\right)  $:%
\[
P\left(  g,N\right)  =\operatorname*{EX}\limits_{X\in\mathcal{D}\left[
n\right]  \left(  g,N\right)  }P\left(  X\right)  .
\]
We then have the following surprising characterization:

\begin{lemma}
\label{univ}For all sufficiently large $n$ and if $T\leq\sqrt{n}/3$, there
exists a bivariate polynomial $q\left(  g,N\right)  $\ of degree at most $2T$
such that if $\left(  g,N\right)  $\ is a quasilattice point, then%
\[
\left|  P\left(  g,N\right)  -q\left(  g,N\right)  \right|  <0.182
\]
(where the constant $0.182$ can be made arbitrarily small by adjusting parameters).
\end{lemma}

\begin{proof}
Let $I$ be a product of $\Delta\left(  x_{i},h\right)  \ $variables, with
degree $r\left(  I\right)  $, and let $I\left(  X\right)  \in\left\{
0,1\right\}  $\ be $I$ evaluated on input $X$. \ Then define%
\[
\gamma\left(  I,g,N\right)  =\operatorname*{EX}\limits_{X\in\mathcal{D}\left[
n\right]  \left(  g,N\right)  }I\left(  X\right)
\]
to be the probability that monomial $I$\ evaluates to $1$ when the input is
drawn from $\mathcal{D}_{n}\left(  g,N\right)  $. \ Then by Lemma \ref{poly},
$P\left(  X\right)  $\ is a polynomial of degree at most $2T$\ over $X$, so%
\[
P\left(  g,N\right)  =\operatorname*{EX}\limits_{X\in\mathcal{D}\left[
n\right]  \left(  g,N\right)  }P\left(  X\right)  =\operatorname*{EX}%
\limits_{X\in\mathcal{D}\left[  n\right]  \left(  g,N\right)  }\sum
_{I:r\left(  I\right)  \leq2t}\beta_{I}I\left(  X\right)  =\sum_{I:r\left(
I\right)  \leq2T}\beta_{I}\gamma\left(  I,g,N\right)
\]
for some coefficients $\beta_{I}$.

We now calculate $\gamma\left(  I,g,N\right)  $. \ Assume without loss of
generality that for all $\Delta\left(  x_{i},h_{1}\right)  ,\Delta\left(
x_{j},h_{2}\right)  \in I$, either $i\neq j$\ or $h_{1}=h_{2}$, since
otherwise $\gamma\left(  I,g,N\right)  =0$.

Define the ``range'' $Z\left(  I\right)  $\ of $I$ to be the set of all $h$
such that $\Delta\left(  x_{i},h\right)  \in I$. \ Let $w\left(  I\right)
=\left|  Z\left(  I\right)  \right|  $; then we write $Z\left(  I\right)
=\left\{  z_{1},\ldots,z_{w\left(  I\right)  }\right\}  $. $\ $Clearly
$\gamma\left(  I,g,N\right)  =0$\ unless $Z\left(  I\right)  \in S$, where $S$
is the range of $\widehat{X}$. \ By assumption,%
\[
\frac{N}{g}\geq\frac{n}{\sqrt{n}}\geq2T\geq r\left(  I\right)
\]
\ so the number of possible $S$ is $\dbinom{n}{N/g}$\ and, of these, the
number that contain $Z$\ is $\dbinom{n-w\left(  I\right)  }{N/g-w\left(
I\right)  }$.

Then, conditioned on $Z\in S$, what is the probability that $\gamma\left(
I,g,N\right)  =1$? \ The total number of $g$-1\ functions with domain size
$N$\ is $N!/\left(  g!\right)  ^{N/g},$ since we can permute the $N$\ function
values arbitrarily, but must not count permutations that act only within the
$N/g$\ constant-value blocks of size $g$.

Among these functions, how many satisfy $\gamma\left(  I,g,N\right)  =1$?
\ Suppose that, for each $1\leq j\leq w\left(  I\right)  $, there are
$r_{j}\left(  I\right)  $\ distinct $i$\ such that $\Delta\left(  x_{i}%
,z_{j}\right)  \in I$. \ Clearly%
\[
r_{1}\left(  I\right)  +\cdots+r_{w\left(  I\right)  }\left(  I\right)
=r\left(  I\right)  .
\]
Then we can permute the $\left(  N-r\left(  I\right)  \right)  !$ function
values outside of $I$ arbitrarily, but must not count permutations that act
only within the $N/g$ constant-value blocks, which have size either $g$ or
$g-r_{i}\left(  I\right)  $ for some $i$. \ So the number of functions for
which $\gamma\left(  I,g,N\right)  =1$\ is%
\[
\frac{\left(  N-r\left(  I\right)  \right)  !}{\left(  g!\right)
^{N/g-w\left(  I\right)  }%
{\displaystyle\prod\nolimits_{i=1}^{w\left(  I\right)  }}
\left(  g-r_{i}\left(  I\right)  \right)  !}.
\]

Putting it all together, \
\begin{align*}
\gamma\left(  I,g,N\right)   &  =\frac{\dbinom{n-w\left(  I\right)
}{N/g-w\left(  I\right)  }}{\dbinom{n}{N/g}}\cdot\frac{\left(  N-r\left(
I\right)  \right)  !\left(  g!\right)  ^{N/g}}{\left(  g!\right)
^{N/g-w\left(  I\right)  }N!%
{\displaystyle\prod\nolimits_{i=1}^{w\left(  I\right)  }}
\left(  g-r_{i}\left(  I\right)  \right)  !}\\
&  =\frac{\left(  N-r\left(  I\right)  \right)  !\left(  n-w\left(  I\right)
\right)  !\left(  N/g\right)  !}{N!n!\left(  N/g-w\left(  I\right)  \right)
!}\cdot\frac{\left(  g!\right)  ^{w\left(  I\right)  }}{%
{\displaystyle\prod\nolimits_{i=1}^{w\left(  I\right)  }}
\left(  g-r_{i}\left(  I\right)  \right)  !}\\
&  =\frac{\left(  N-r\left(  I\right)  \right)  !}{N!}\frac{\left(  n-w\left(
I\right)  \right)  !}{n!}\cdot%
{\displaystyle\prod\limits_{i=0}^{w\left(  I\right)  -1}}
\left(  \frac{N}{g}-i\right)
{\displaystyle\prod\limits_{i=1}^{w\left(  I\right)  }}
\left[  g%
{\displaystyle\prod\limits_{j=1}^{r\left[  i\right]  \left(  I\right)  -1}}
\left(  g-j\right)  \right] \\
&  =\frac{\left(  N-2T\right)  !}{N!}\frac{\left(  n-w\left(  I\right)
\right)  !}{n!}\cdot%
{\displaystyle\prod\limits_{i=r\left(  I\right)  }^{2T-1}}
\left(  N-i\right)
{\displaystyle\prod\limits_{i=0}^{w\left(  I\right)  -1}}
\left(  N-gi\right)
{\displaystyle\prod\limits_{i=1}^{w\left(  I\right)  }}
{\displaystyle\prod\limits_{j=1}^{r\left[  i\right]  \left(  I\right)  -1}}
\left(  g-j\right) \\
&  =\frac{\left(  N-2T\right)  !n!}{N!\left(  n-2T\right)  !}\widetilde
{q}_{n,T,I}\left(  g,N\right)
\end{align*}
where%
\[
\widetilde{q}_{n,T,I}\left(  g,N\right)  =\frac{\left(  n-w\left(  I\right)
\right)  !\left(  n-2T\right)  !}{\left(  n!\right)  ^{2}}\cdot%
{\displaystyle\prod\limits_{i=r\left(  I\right)  }^{2T-1}}
\left(  N-i\right)
{\displaystyle\prod\limits_{i=0}^{w\left(  I\right)  -1}}
\left(  N-gi\right)
{\displaystyle\prod\limits_{i=1}^{w\left(  I\right)  }}
{\displaystyle\prod\limits_{j=1}^{r\left[  i\right]  \left(  I\right)  -1}}
\left(  g-j\right)
\]
\ is a bivariate polynomial of total degree at most%
\[
\left(  2T-r\left(  I\right)  \right)  +w\left(  I\right)  +\left(  r\left(
I\right)  -w\left(  I\right)  \right)  =2T.
\]
(Note that in the case $r_{i}\left(  I\right)  >g$ for some $i$, this
polynomial evaluates to $0$, which is what it ought to do.) \ Hence%
\[
P\left(  g,N\right)  =\sum_{I:r\left(  I\right)  \leq2T}\beta_{I}\gamma\left(
I,g,N\right)  =\frac{\left(  N-2T\right)  !n!}{N!\left(  n-2T\right)
!}q\left(  g,N\right)
\]
where%
\[
q\left(  g,N\right)  =\sum_{I:r\left(  I\right)  \leq2T}\beta_{I}\widetilde
{q}_{n,T,I}\left(  g,N\right)  .
\]

Clearly%
\[
\frac{\left(  N-2T\right)  !n!}{N!\left(  n-2T\right)  !}\leq1.
\]
Since $N\leq n+n/\left(  10T\right)  $ and $T\leq\sqrt{n}/3$, we also have%
\begin{align*}
\frac{\left(  N-2T\right)  !n!}{N!\left(  n-2T\right)  !}  &  \geq\left(
\frac{n-2T+1}{N-2T+1}\right)  ^{2T}\\
&  \geq\left(  1-\frac{n/\left(  10T\right)  }{N-2T+1}\right)  ^{2T}\\
&  \geq\left(  1-\frac{n}{10\left[  n-\left(  2T+1\right)  /n\right]
}\frac{1}{T}\right)  ^{2T}\\
&  \geq\exp\left\{  -\frac{1}{5}\frac{n}{n-\left(  2T+1\right)  /n}\right\} \\
&  \geq0.818
\end{align*}
for all sufficiently large $n$.

Thus, since $0\leq P\left(  g,N\right)  \leq1$,%
\[
\left|  P\left(  g,N\right)  -q\left(  g,N\right)  \right|  <0.182
\]
and we are done.
\end{proof}

\section{Lower Bound\label{lb}}

We are now ready to prove a lower bound for the collision problem. \ To do so,
we generalize an approximation theory result due to Rivlin and Cheney
\cite{rc}\ and (independently) Ehlich and Zeller \cite{ez}. \ That result was
applied to query complexity by Nisan and Szegedy \cite{ns}\ and later by Beals
et al. \cite{bbcmw}.

\begin{theorem}
\label{thetheorem}$Q_{2}\left(  \operatorname*{Col}_{n}\right)  =\Omega\left(
n^{1/5}\right)  .$
\end{theorem}

\begin{proof}
Let $g$\ have range $1\leq g\leq G$. \ Then the quasilattice points $\left(
g,N\right)  $\ all lie in the rectangular region $R=\left[  1,G\right]
\times\left[  n,n+n/\left(  10T\right)  \right]  $. \ Recalling the polynomial
$q\left(  g,N\right)  $\ from Lemma \ref{univ},\ define%
\[
d\left(  q\right)  =\max_{\left(  g,N\right)  \in R}\left(  \max\left\{
\left|  \frac{\partial q}{\partial g}\right|  ,\frac{n}{10T\left(  G-1\right)
}\cdot\left|  \frac{\partial q}{\partial N}\right|  \right\}  \right)  .
\]
Suppose without loss of generality that we require $P\left(  1,n\right)
\leq1/10$\ and $P\left(  2,n\right)  \geq9/10$\ (that is, algorithm $A$
distinguishes 1-1 from 2-1 functions with error probability at most $1/10$).
\ Then, since%
\[
\left|  P\left(  g,N\right)  -q\left(  g,N\right)  \right|  <0.182
\]
by the Intermediate Value Theorem we have%
\[
d\left(  q\right)  \geq\max_{1\leq g\leq2}\frac{\partial q}{\partial
g}>0.8-2\left(  0.182\right)  =\allowbreak0.436.
\]
$\,$

\ An inequality due to Markov (see \cite{cheney,ns})\ states that, for a
univariate polynomial $p$, if $b_{1}\leq p\left(  x\right)  \leq b_{2}$\ for
all\ $a_{1}\leq x\leq a_{2}$, then%
\[
\max_{a\left[  1\right]  \leq x\leq a\left[  2\right]  }\left|
\frac{dp\left(  x\right)  }{dx}\right|  \leq\frac{b_{2}-b_{1}}{a_{2}-a_{1}%
}\deg\left(  p\right)  ^{2}.
\]
Clearly for every point $\left(  \widehat{g},\widehat{N}\right)  \in R$, there
exists a quasilattice point $\left(  g,N\right)  $\ for which $\left|
g-\widehat{g}\right|  \leq1$\ and $\left|  N-\widehat{N}\right|  \leq G$.
\ For take $g=\left\lceil \widehat{g}\right\rceil $---or, in the special case
$\widehat{g}=1$, take $g=2$, since there is only one quasilattice point with
$g=1$.

Furthermore, since $P\left(  g,N\right)  $\ represents an acceptance
probability at such a point, we have%
\[
-0.182<q\left(  g,N\right)  <1.182.
\]
Observe that for all $\left(  \widehat{g},\widehat{N}\right)  \in R$, \
\[
-0.182-\left(  \frac{10TG\left(  G-1\right)  }{n}+1\right)  d\left(  q\right)
<q\left(  \widehat{g},\widehat{N}\right)  <1.182+\left(  \frac{10TG\left(
G-1\right)  }{n}+1\right)  d\left(  q\right)  .
\]
For consider a quasilattice point close to $\left(  \widehat{g},\widehat
{N}\right)  $, and note that the maximum-magnitude derivative is at most
$d\left(  q\right)  $\ in the $g$ direction and $10T\left(  G-1\right)
d\left(  q\right)  /n$ in the $N$ direction.

Let $\left(  g^{\ast},N^{\ast}\right)  $\ be a point in $R$ at which the
weighted maximum-magnitude derivative $d\left(  q\right)  $\ is attained.
\ Suppose first that the maximum is attained in the $g$\ direction. \ Then
$q\left(  g,N^{\ast}\right)  $\ (with $N^{\ast}$ constant) is a univariate
polynomial with%
\[
\left|  \frac{dq\left(  g,N^{\ast}\right)  }{dg}\right|  >0.436
\]
for some $1\leq g\leq G$. \ So%
\begin{align*}
2T  &  \geq\deg\left(  q\left(  g,N\right)  \right) \\
&  \geq\deg\left(  q\left(  g,N^{\ast}\right)  \right) \\
&  \geq\sqrt{\frac{d\left(  q\right)  \left(  G-1\right)  }{1.364+2d\left(
q\right)  \left(  1+10TG\left(  G-1\right)  /n\right)  }}\\
&  \geq\sqrt{\frac{0.436\left(  G-1\right)  n}{2.236n+8.720TG\left(
G-1\right)  }}\\
&  =\Omega\left(  \min\left\{  \sqrt{G},\sqrt{\frac{n}{TG}}\right\}  \right)
.
\end{align*}

Similarly, suppose the maximum $d\left(  q\right)  $\ is attained\ in the $N$
direction. \ Then $q\left(  g^{\ast},N\right)  $\ (with $g^{\ast}$ constant)
is a univariate polynomial with%
\[
\left|  \frac{dq\left(  g^{\ast},N\right)  }{dN}\right|  >\frac{0.436T\left(
G-1\right)  }{n}%
\]
for some $n\leq N\leq n+n/\left(  10T\right)  $. \ So%
\[
2T\geq\sqrt{\frac{\left(  10T\left(  G-1\right)  /n\right)  d\left(  q\right)
n/\left(  10T\right)  }{1.364+2d\left(  q\right)  \left(  1+10TG\left(
G-1\right)  /n\right)  }}\geq\Omega\left(  \min\left\{  \sqrt{G}%
,\sqrt{\frac{n}{TG}}\right\}  \right)  .
\]

One can show that the lower bound on $T$\ is optimized when we take
$G=n^{2/5}\leq\sqrt{n}$. \ Then%
\begin{align*}
T  &  =\Omega\left(  \min\left\{  n^{1/5},\frac{\sqrt{n}}{\sqrt{T}n^{1/5}%
}\right\}  \right)  ,\\
T  &  =\Omega\left(  n^{1/5}\right)
\end{align*}
and we are done.
\end{proof}

\section{Acknowledgments}

I am grateful to Yaoyun Shi, Ronald de Wolf, Umesh Vazirani, Ashwin Nayak, and
Andris Ambainis for helpful comments; to Leonard Schulman, Lawrence Ip, Jordan
Kerenidis, and John Preskill for discussions during earlier stages of this
work; and to James Lee, Alex Halderman, and Elham Kashefi for discussions and
references regarding Section \ref{motivation}.

\section{Appendix: Set Comparison\label{setcomp}}

Here we show that $Q_{2}\left(  \operatorname*{SetComp}_{n}\right)
=\Omega\left(  n^{1/7}\right)  $, where $\operatorname*{SetComp}_{n}$\ is the
set comparison problem of size $n$ as defined in Section \ref{erasure}. \ We
need only redo the proof of Lemma \ref{univ}; then Theorem \ref{thetheorem}%
\ goes through largely unchanged.

The idea is the following. \ We need a distribution of inputs with a parameter
$g$, such that the inputs are one-to-one when $g=1$ or $g=2$---since otherwise
the problem of distinguishing $g=1$\ from $g=2$\ would be ill-defined for
erasing oracles. \ On the other hand, the inputs must \textit{not} be
one-to-one for all $g>2$---since otherwise the lower bound for standard
oracles would apply also to erasing oracles, and we could not obtain a
separation between the two. \ Finally, the algorithm's acceptance probability
must be close to a polynomial in $g$.

Our solution is to consider $\kappa\left(  g\right)  $-to-one inputs, where%

\[
\kappa\left(  g\right)  =4g^{2}-12g+9.
\]
is a quadratic with $\kappa\left(  1\right)  =\kappa\left(  2\right)  =1$.
\ The total range of the inputs (on sequences $X$ and $Y$\ combined) has size
roughly $n/g$; thus, we can tell the $g=1$\ inputs apart from the
$g=2$\ inputs using an erasing oracle, even though $\kappa\left(  g\right)
$\ is the same for both. \ The disadvantage is that, because $\kappa\left(
g\right)  $\ increases quadratically rather than linearly in $g$, the
quasilattice points become sparse more quickly. \ That is what weakens the
lower bound from $\Omega\left(  n^{1/5}\right)  $\ to $\Omega\left(
n^{1/7}\right)  $. \ We note that, using the ideas of Shi \cite{shi}, one can
improve our lower bound on $Q_{2}\left(  \operatorname*{SetComp}_{n}\right)
$\ to $\Omega\left(  n^{1/6}\right)  $.

Call $\left(  g,N,M\right)  \in\Re^{3}$\ an $\left(  n,T\right)
$\textit{-super-quasilattice point} if and only if

\begin{enumerate}
\item[(1)] $g$ is an integer in $\left[  1,n^{1/3}\right]  $,

\item[(2)] $N$ and $M$ are integers in $\left[  n,n\left(  1+1/\left(
100T\right)  \right)  \right]  $,

\item[(3)] $g$ divides $N$,

\item[(4)] if $g=1$\ then $N=n$,

\item[(5)] $\kappa\left(  g\right)  $ divides $M$, and

\item[(6)] if $g=2$ then $M=n$.
\end{enumerate}

For super-quasilattice point $\left(  g,N,M\right)  $, we draw input $\left(
X,Y\right)  =\left(  x_{1}\ldots x_{n},y_{1}\ldots y_{n}\right)  $\ from
distribution $\mathcal{L}_{n}\left(  g,N,M\right)  $\ as follows. \ We first
choose a set $S\subseteq\left\{  1,\ldots,2n\right\}  $\ with $\left|
S\right|  =2N/g\leq2n$ uniformly at random. \ We then choose two sets
$S_{X},S_{Y}\subseteq S$\ with $\left|  S_{X}\right|  =\left|  S_{X}\right|
=M/\kappa\left(  g\right)  \leq\left|  S\right|  $, uniformly at random and
independently. \ Next we choose $\kappa\left(  g\right)  $-1 functions
$\widehat{X}=\widehat{x}_{1}\ldots\widehat{x}_{N}$ $:\left\{  1,\ldots
,M\right\}  \rightarrow S_{X}$ and $\widehat{Y}=\widehat{y}_{1}\ldots
\widehat{y}_{N}$ $:\left\{  1,\ldots,M\right\}  \rightarrow S_{Y}$\ uniformly
at random and independently. \ Finally we let $x_{i}=\widehat{x}_{i}$\ and
$y_{i}=\widehat{y}_{i}$\ for each $1\leq i\leq n$.

Define sets $X_{S}=\left\{  x_{1},\ldots,x_{n}\right\}  $\ and $Y_{S}=\left\{
y_{1},\ldots,y_{n}\right\}  $. \ Suppose $g=1$ and $N=M=n$; then by Chernoff
bounds,%
\[
\Pr_{\left(  X,Y\right)  \in\mathcal{L}\left[  n\right]  \left(  1,n,n\right)
}\left[  \left|  X_{S}\cup Y_{S}\right|  <1.1n\right]  \leq2e^{-n/10}.
\]
Thus, if algorithm $A$ can distinguish $\left|  X_{S}\cup Y_{S}\right|
=n$\ from $\left|  X_{S}\cup Y_{S}\right|  \geq1.1n$\ with probability at
least $9/10$, then it can distinguish $\left(  X,Y\right)  \in\mathcal{L}%
_{n}\left(  1,n,n\right)  $\ from $\left(  X,Y\right)  \in\mathcal{L}%
_{n}\left(  2,n,n\right)  $\ with probability at least $9/10-2e^{-n/10}$. \ So
a lower bound for the latter problem implies an equivalent lower bound for the former.

Define $P\left(  X,Y\right)  $\ to be the probability that the algorithm
returns that $X$ and $Y$ are far on input $\left(  X,Y\right)  $, and let%
\[
P\left(  g,N,M\right)  =\operatorname*{EX}\limits_{\left(  X,Y\right)
\in\mathcal{L}\left[  n\right]  \left(  g,N,M\right)  }P\left(  X,Y\right)  .
\]
We then have

\begin{lemma}
\label{univ2}For all sufficiently large $n$ and if $T\leq n^{1/3}/8$, there
exists a trivariate polynomial $q\left(  g,N,M\right)  $\ of degree at most
$8T$ such that if $\left(  g,N,M\right)  $\ is a super-quasilattice point,
then%
\[
\left|  P\left(  g,N,M\right)  -q\left(  g,N,M\right)  \right|  <\varepsilon
\]
for some constant $0<\varepsilon<1/2$.
\end{lemma}

\begin{proof}
By analogy to Lemma \ref{poly}, $P\left(  X,Y\right)  $\ is a multilinear
polynomial of degree at most $2T$\ over variables of the form $\Delta\left(
x_{i},h\right)  $\ and $\Delta\left(  y_{i},h\right)  $. \ Let $I\left(
X,Y\right)  =I_{X}\left(  X\right)  I_{Y}\left(  Y\right)  $ where $I_{X}%
$\ is\ a product of $r_{X}\left(  I\right)  $\ distinct $\Delta\left(
x_{i},h\right)  \ $variables and $I_{Y}$ is a product of $r_{Y}\left(
I\right)  \ $distinct$\ \Delta\left(  y_{i},h\right)  $\ variables. \ Let
$r\left(  I\right)  =r_{X}\left(  I\right)  +r_{Y}\left(  I\right)  $.
\ Define%
\[
\gamma\left(  I,g,N,M\right)  =\operatorname*{EX}\limits_{\left(  X,Y\right)
\in\mathcal{L}\left[  n\right]  \left(  g,N,M\right)  }I\left(  X,Y\right)  ;
\]
then%
\[
P\left(  g,N,M\right)  =\sum_{I:r\left(  I\right)  \leq2T}\beta_{I}%
\gamma\left(  I,g,N,M\right)
\]
for some coefficients $\beta_{I}$.

We now calculate $\gamma\left(  I,g,N,M\right)  $. \ As before we assume there
are no pairs of variables $\Delta\left(  x_{i},h_{1}\right)  ,\Delta\left(
x_{i},h_{2}\right)  \in I$ with $h_{1}\neq h_{2}$.

Let $Z_{X}\left(  I\right)  $\ be the range of $I_{X}$ and let $Z_{Y}\left(
I\right)  $\ be the range of $I_{Y}$.\ \ Then let $Z\left(  I\right)
=Z_{X}\left(  I\right)  \cup Z_{Y}\left(  I\right)  $. \ Let $w_{X}\left(
I\right)  =\left|  Z_{X}\left(  I\right)  \right|  $, $w_{Y}\left(  I\right)
=\left|  Z_{Y}\left(  I\right)  \right|  $, and $w\left(  I\right)  =\left|
Z\left(  I\right)  \right|  $. \ By assumption%
\[
\frac{N}{g}\geq\frac{M}{\kappa\left(  g\right)  }\geq\frac{1}{4}n^{1/3}\geq2T
\]
so%
\[
\Pr\left[  Z\left(  I\right)  \subseteq S\right]  =\frac{\dbinom{2n-w\left(
I\right)  }{2N/g-w\left(  I\right)  }}{\dbinom{2n}{2N/g}}.
\]
Then the probability that $Z_{X}\left(  I\right)  \subseteq S_{X}$ given
$Z\left(  I\right)  \subseteq S$ is%
\[
\frac{\dbinom{2N/g-w_{X}\left(  I\right)  }{M/\kappa\left(  g\right)
-w_{X}\left(  I\right)  }}{\dbinom{2N/g}{M/\kappa\left(  g\right)  }}%
\]
and similarly for the probability that $Z_{Y}\left(  I\right)  \subseteq
S_{Y}$ given $Z\left(  I\right)  \subseteq S$.

Let $r_{X,1}\left(  I\right)  ,\ldots,r_{X,w\left[  X\right]  \left(
I\right)  }\left(  I\right)  $\ be the multiplicities of the range elements in
$Z_{X}\left(  I\right)  $, so that $r_{X,1}\left(  I\right)  +\cdots
+r_{X,w\left[  X\right]  \left(  I\right)  }\left(  I\right)  =r_{X}\left(
I\right)  $. \ Then%
\[
\Pr\left[  I_{X}\left(  X\right)  \,\,|\,\,Z_{X}\left(  I\right)  \subseteq
S_{X}\right]  =\frac{\left(  M-r_{X}\left(  I\right)  \right)  !}{M!}%
{\displaystyle\prod\limits_{i=1}^{w\left[  X\right]  \left(  I\right)  }}
{\displaystyle\prod\limits_{j=0}^{r\left[  X,i\right]  \left(  I\right)  -1}}
\left(  \kappa\left(  g\right)  -j\right)
\]
and similarly for $\Pr\left[  I_{Y}\left(  Y\right)  \,\,|\,\,Z_{Y}\left(
I\right)  \subseteq S_{Y}\right]  $.

Putting it all together,%
\begin{align*}
\gamma\left(  I,g,N,M\right)  =  &  \frac{\dbinom{2n-w\left(  I\right)
}{2N/g-w\left(  I\right)  }}{\dbinom{2n}{2N/g}}\frac{\left(  M-r_{X}\left(
I\right)  \right)  !}{M!}\frac{\left(  M-r_{Y}\left(  I\right)  \right)
!}{M!}\frac{\dbinom{2N/g-w_{X}\left(  I\right)  }{M/\kappa\left(  g\right)
-w_{X}\left(  I\right)  }}{\dbinom{2N/g}{M/\kappa\left(  g\right)  }}\times\\
&
{\displaystyle\prod\limits_{i=1}^{w\left[  X\right]  \left(  I\right)  }}
{\displaystyle\prod\limits_{j=0}^{r\left[  X,i\right]  \left(  I\right)  -1}}
\left(  \kappa\left(  g\right)  -j\right)  \frac{\dbinom{2N/g-w_{Y}\left(
I\right)  }{M/\kappa\left(  g\right)  -w_{Y}\left(  I\right)  }}{\dbinom
{2N/g}{M/\kappa\left(  g\right)  }}%
{\displaystyle\prod\limits_{i=1}^{w\left[  Y\right]  \left(  I\right)  }}
{\displaystyle\prod\limits_{j=0}^{r\left[  Y,i\right]  \left(  I\right)  -1}}
\left(  \kappa\left(  g\right)  -j\right) \\
=  &  \frac{\left(  2n-w\left(  I\right)  \right)  !}{\left(  2n\right)
!}\frac{\left(  M-r_{X}\left(  I\right)  \right)  !}{M!}\frac{\left(
M-r_{Y}\left(  I\right)  \right)  !}{M!}\frac{\left(  2N/g-w_{X}\left(
I\right)  \right)  !}{\left(  2N/g-w\left(  I\right)  \right)  !}\frac{\left(
2N/g-w_{Y}\left(  I\right)  \right)  !}{\left(  2N/g\right)  !}\theta
_{I}\left(  g,M\right)
\end{align*}
where%
\begin{align*}
&  \theta_{I}\left(  g,M\right)  =\\
&
{\displaystyle\prod\limits_{i=0}^{w\left[  X\right]  \left(  I\right)  -1}}
\left(  M-i\kappa\left(  g\right)  \right)
{\displaystyle\prod\limits_{i=1}^{w\left[  X\right]  \left(  I\right)  }}
{\displaystyle\prod\limits_{j=1}^{r\left[  X,i\right]  \left(  I\right)  -1}}
\left(  \kappa\left(  g\right)  -j\right)
{\displaystyle\prod\limits_{i=0}^{w\left[  Y\right]  \left(  I\right)  -1}}
\left(  M-i\kappa\left(  g\right)  \right)
{\displaystyle\prod\limits_{i=1}^{w\left[  Y\right]  \left(  I\right)  }}
{\displaystyle\prod\limits_{j=1}^{r\left[  Y,i\right]  \left(  I\right)  -1}}
\left(  \kappa\left(  g\right)  -j\right)
\end{align*}
is a bivariate polynomial in $\left(  g,M\right)  $\ of total degree at most
$2r\left(  I\right)  $.

Thus%
\begin{align*}
\gamma\left(  I,g,N,M\right)  =  &  \frac{\left(  2n-w\left(  I\right)
\right)  !}{\left(  2n\right)  !}\left[  \frac{\left(  M-2T\right)
!n!}{M!\left(  n-2T\right)  !}\right]  ^{2}\left(  \frac{\left(  n-2T\right)
!}{n!}\right)  ^{2}\prod_{i=r\left[  X\right]  \left(  I\right)  }%
^{2T-1}\left(  M-i\right)  \prod_{i=r\left[  Y\right]  \left(  I\right)
}^{2T-1}\left(  M-i\right)  \times\\
&  \frac{\left(  2N/g-w_{X}\left(  I\right)  \right)  \cdots\left(
2N/g-\left(  w\left(  I\right)  -1\right)  \right)  }{\left(  2N/g\right)
\left(  2N/g-1\right)  \cdots\left(  2N/g-\left(  w_{Y}\left(  I\right)
-1\right)  \right)  }\theta_{I}\left(  g,M\right) \\
=  &  \frac{\left(  2n\right)  ^{2T}}{\left(  2N\right)  \left(  2N-g\right)
\cdots\left(  2N-\left(  2T-1\right)  g\right)  }\left[  \frac{\left(
M-2T\right)  !n!}{M!\left(  n-2T\right)  !}\right]  ^{2}\widetilde{q}%
_{n,T,I}\left(  g,N,M\right)
\end{align*}
where%
\begin{align*}
\widetilde{q}_{n,T,I}\left(  g,N,M\right)  =  &  \frac{\left(  2n-w\left(
I\right)  \right)  !}{\left(  2n\right)  !\left(  2n\right)  ^{2T}}\left(
\frac{\left(  n-2T\right)  !}{n!}\right)  ^{2}g^{w\left[  X\right]  \left(
I\right)  +w\left[  Y\right]  \left(  I\right)  -w\left(  I\right)  }%
\theta_{I}\left(  g,M\right)  \prod_{i=r\left[  X\right]  \left(  I\right)
}^{2T-1}\left(  M-i\right)  \times\\
&  \prod_{i=r\left[  Y\right]  \left(  I\right)  }^{2T-1}\left(  M-i\right)
\prod_{i=w\left[  X\right]  \left(  I\right)  }^{w\left(  I\right)  -1}\left(
2N-ig\right)  \prod_{i=w\left[  Y\right]  \left(  I\right)  }^{2T-1}\left(
2N-ig\right)
\end{align*}
is a trivariate polynomial in $\left(  g,N,M\right)  $ of total degree at most%
\[
\left(  4T-r\left(  I\right)  \right)  +2r\left(  I\right)  +\left(
w_{X}\left(  I\right)  +w_{Y}\left(  I\right)  -w\left(  I\right)  \right)
+\left(  w\left(  I\right)  -w_{X}\left(  I\right)  \right)  +\left(
2T-w_{Y}\left(  I\right)  \right)  \leq8T.
\]

Thus%
\[
P\left(  g,N,M\right)  =\frac{\left(  2n\right)  ^{2T}}{%
{\displaystyle\prod\limits_{i=0}^{2T-1}}
\left(  2N-gi\right)  }\left[  \frac{\left(  M-2T\right)  !n!}{M!\left(
n-2T\right)  !}\right]  ^{2}q\left(  g,N,M\right)
\]
where $q\left(  g,N,M\right)  $\ is a polynomial of total degree at most $8T$.
\ The argument that $q$\ approximates $P$\ to within a constant is analogous
to that of Lemma \ref{univ}; note that%
\[
\left|  \frac{\left(  2n\right)  ^{2T}}{%
{\displaystyle\prod\limits_{i=0}^{2T-1}}
\left(  2N-gi\right)  }-1\right|  =O\left[  \left(  1+\frac{1}{T}+\frac{gT}%
{n}\right)  ^{2T}\right]  =O\left(  1\right)
\]
since $g\leq n^{1/3}\ $and $T\leq n^{1/3}/8$.
\end{proof}

\begin{theorem}
\label{thetheorem2}$Q_{2}\left(  \operatorname*{SetComp}_{n}\right)
=\Omega\left(  n^{1/7}\right)  $.
\end{theorem}

\begin{proof}
[Proof sketch]The proof is analogous to that of Theorem \ref{thetheorem}.
\ Let $g\in\left[  1,G\right]  $ for some $G\leq n^{1/3}$. \ Then the
super-quasilattice points $\left(  g,N,M\right)  $\ all lie in $R=\left[
1,G\right]  \times\left[  n,n+n/\left(  100T\right)  \right]  ^{2}$. \ Define
$d(q)$ to be%
\[
\max_{\left(  g,N,M\right)  \in R}\left(  \max\left\{  \left|  \frac{\partial
q}{\partial g}\right|  ,\frac{n/100T}{\left(  G-1\right)  }\left|
\frac{\partial q}{\partial N}\right|  ,\frac{n/100T}{\left(  G-1\right)
}\left|  \frac{\partial q}{\partial M}\right|  \right\}  \right)  .
\]
Then $d\left(  q\right)  \geq\delta$ for some constant $\delta>0$, by Lemma
\ref{univ2}.

For every point $\left(  \widehat{g},\widehat{N},\widehat{M}\right)  \in R$,
there exists a super-quasilattice point $\left(  g,N,M\right)  $\ such that
$\left|  g-\widehat{g}\right|  \leq1$, $\left|  N-\widehat{N}\right|  \leq G$,
and $\left|  M-\widehat{M}\right|  \leq\kappa\left(  G\right)  $.\ \ Hence,
$q\left(  \widehat{g},\widehat{N},\widehat{M}\right)  $\ can deviate from
$\left[  0,1\right]  $\ by at most%
\[
O\left(  \left(  \frac{TG^{3}}{n}+1\right)  d\left(  q\right)  \right)  .
\]

Let $\left(  g^{\ast},N^{\ast},M^{\ast}\right)  $\ be a point in $R$ at which
$d\left(  q\right)  $\ is attained. \ Suppose $d\left(  q\right)  $\ is
attained in the $g$ direction; the cases of the $N$ and $M$ directions are
analogous. \ Then $q\left(  g,N^{\ast},M^{\ast}\right)  $\ is a univariate
polynomial in $g$, and%
\begin{align*}
8T  &  \geq\deg\left(  q\left(  g,N^{\ast},M^{\ast}\right)  \right) \\
&  =\Omega\left(  \sqrt{\frac{d\left(  q\right)  G}{1+d\left(  q\right)
+d\left(  q\right)  TG^{3}/n}}\right) \\
&  =\Omega\left(  \min\left\{  \sqrt{G},\sqrt{\frac{n}{TG^{2}}}\right\}
\right)  .
\end{align*}

One can show that the bound is optimized when we take $G=n^{2/7}\leq n^{1/3}$.
\ Then%
\begin{align*}
T  &  =\Omega\left(  \min\left\{  n^{1/7},\frac{\sqrt{n}}{\sqrt{T}n^{2/7}%
}\right\}  \right)  ,\\
T  &  =\Omega\left(  n^{1/7}\right)  .
\end{align*}
\end{proof}

\begin{thebibliography}{9}                                                                                                %

\bibitem {ambainis}A. Ambainis. \ Quantum lower bounds by quantum arguments.
\ \textit{Proceedings of STOC'2000}, pages 636--643, 2000. \ Journal version
to appear in \textit{Journal of Computer and System Sciences}.
\ quant-ph/0002066\footnote{Available at \texttt{www.arxiv.org}.}.

\bibitem {bbcmw}R. Beals, H. Buhrman, R. Cleve, M. Mosca, and R. de Wolf.
\ Quantum lower bounds by polynomials. \ \textit{Proceedings of FOCS'98},
pages 352--361, 1998. \ quant-ph/9802049.

\bibitem {bbbv}C. Bennett, E. Bernstein, G. Brassard, and U. Vazirani.
\ Strengths and weaknesses of quantum computing. \ \textit{SIAM Journal on
Computing}, 26(5):1510--1523, 1997. \ quant-ph/9701001.

\bibitem {bv}E. Bernstein and U. Vazirani. \ Quantum complexity theory.
\ \textit{SIAM\ Journal on Computing}, 26(5):1411--1473, 1997.

\bibitem {bht}G. Brassard, P. H\o yer, and A. Tapp. \ Quantum algorithm for
the collision problem. \ \textit{ACM\ SIGACT\ News (Cryptology Column)},
28:14--19, 1997. \ quant-ph/9705002.

\bibitem {bdhhmsw}H. Buhrman, C. D\"{u}rr, M. Heiligman, P. H\o yer, F.
Magniez, M. Santha, and R. de Wolf. \ Quantum algorithms for element
distinctness. \ \textit{Proceedings of IEEE\ Conference on Computational
Complexity (CCC'2001)}, pages 131--137, 2001. \ quant-ph/0007016.

\bibitem {bsp}S. Bakhtiari, R. Safavi-Naini, and J. Pieprzyk. \ Cryptographic
hash functions: a survey. \ Technical Report 95-09, Department of Computer
Science, University of Wollongong, July 1995. \ Available at ftp://ftp.cs.uow.edu.au/pub/papers/1995/tr-95-09.ps.Z.

\bibitem {cheney}E. W. Cheney. \ \textit{Introduction to approximation
theory}, McGraw-Hill, 1966.

\bibitem {damgard}I. B. Damg\aa rd. \ Collision free hash functions and public
key signature schemes. \ \textit{Proceedings of Eurocrypt'87}, Volume 304 of
\textit{Lecture Notes in Computer Science} (Springer-Verlag), 1988.

\bibitem {eh}M. Ettinger and P. H\o yer. \ On quantum algorithms for
noncommutative hidden subgroups. \ \textit{Advances in Applied Mathematics},
25(3):239--251, 2000.

\bibitem {ez}H. Ehlich and K. Zeller. \ Schwankung von Polynomen zwischen
Gitterpunkten. \ \textit{Mathematische Zeitschrift}, 86:41--44, 1964.

\bibitem {gsvv}M. Grigni, L. Schulman, M. Vazirani, and U. Vazirani. \ Quantum
mechanical algorithms for the nonabelian hidden subgroup problem.
\ \textit{Proceedings of STOC'2001}, pages 68--74, 2001. \ 

\bibitem {grover}L. K. Grover. \ A fast quantum mechanical algorithm for
database search. \ \textit{Proceedings of STOC'96}, pages 212--219, 1996. \ quant-ph/9605043.

\bibitem {kashefi}E. Kashefi, A. Kent, V. Vedral, and K. Banaszek. \ On the
power of quantum oracles, 2001. \ quant-ph/0109104.

\bibitem {mp}M. Minsky and S. Papert. \ \textit{Perceptrons}, MIT\ Press,
1988. \ First appeared in 1968.

\bibitem {nisan}N. Nisan. \ CREW PRAMs and decision trees. \ \textit{SIAM
Journal on Computing}, 20(6):999-1007, 1991.

\bibitem {ns}N. Nisan and M. Szegedy. \ On the degree of Boolean functions as
real polynomials. \ \textit{Computational Complexity}, 4:301--313, 1994.

\bibitem {rains}E. Rains. \ Talk given at AT\&T, Murray Hill, New Jersey, on
March 12, 1997.

\bibitem {rc}T. J. Rivlin and E. W. Cheney. \ A comparison of Uniform
Approximations on an interval and a finite subset thereof. \ \textit{SIAM
Journal on Numerical Analysis}, 3(2):311--320, 1966.

\bibitem {shi}Y. Shi. \ Improving the lower bound on the collision problem to
$n^{1/4}$. \ Manuscript, 2001.

\bibitem {shor}P. Shor. \ Polynomial-time algorithms for prime factorization
and discrete logarithms on a quantum computer. \ \textit{SIAM Journal on
Computing}, 26(5):1484--1509, 1997. \ quant-ph/9508027.

\bibitem {simon}D. Simon. \ On the power of quantum computation.
\ \textit{Proceedings of FOCS'94}, pages 116--123, 1994.

\bibitem {watrous}J. Watrous. \ Succinct quantum proofs for properties of
finite groups. \ \textit{Proceedings of FOCS'2000}, pages 537--546, 2000. \ cs.CC/0009002.\pagebreak 
\end{thebibliography}
\end{document}